\documentclass[Journal,InsideFigs,NoLineNumbers]{ascelike}
\usepackage{graphicx}
\usepackage{natbib}
\usepackage{subfigure}
\usepackage{amsmath}
\usepackage{amsfonts}
\usepackage{amssymb}
\usepackage{amsbsy}
\usepackage{times}
\usepackage{color}
\newcommand\Fr{\mbox{\textit{Fr}}}  

\NameTag{Valiani and Caleffi, Sept. 9, 2015}
\begin{document}
%
\title{FREE SURFACE AXIALLY SYMMETRIC FLOWS \\ AND RADIAL HYDRAULIC JUMPS}
\author{
Alessandro Valiani %
%
\thanks{
Dept.\ of Engineering,
University of Ferrara, 
via G.\ Saragat 1,
Ferrara, ITALY I-44122. E-mail: alessandro.valiani@unife.it}
%
\\
 Valerio Caleffi 
\thanks{
Dept.\ of Engineering,
University of Ferrara, 
via G.\ Saragat 1,
Ferrara, ITALY I-44122. E-mail: valerio.caleffi@unife.it\newline\newline
This material may be downloaded for personal use only. Any other use requires prior permission of the American Society of Civil Engineers. This material may be found at https://doi.org/10.1061/(ASCE)HY.1943-7900.0001104.}
%
\\
%
%
%
%
}
\maketitle
\begin{abstract}
Free surface, axially symmetric shallow flow is analysed in both the centrifugal and centripetal directions.
Referring to the inviscid steady flow over a flat plate characterised by a unique value of specific energy, the analytical sub- and supercritical solutions are determined. Furthermore, the sub- and supercritical steady solutions for the flow with friction over a flat plate are determined, provided that inertial terms are important compared with frictional terms. As the inviscid case, the sub- and supercritical solutions over a bottom topography are determined, provided that the bottom unevenness is compatible with a continuous solution. In the fundamental case of inviscid flow over a flat plate, the discontinuous solution is also analysed for a direct hydraulic jump imposed by proper boundary conditions. In the simple scheme of an inviscid shock of zero length, the jump position and the sequent depths are analytically derived, thus indicating that all results are uniquely functions of one dimensionless number, i. e. the specific energy ratio between the sub- and supercritical flows.
\end{abstract}
%
%
\section{Introduction}
This work is part of a series of research on radial, axially symmetric free surface flows studied for both their intrinsic interest and their significance as test cases to validate numerical models that integrate two-dimensional (2D) Shallow-Water Equations (SWE) \citep{VC2003, CV2012}. Basic contributions typically address the hydraulic jump in radial flow and/or diverging or converging channels, and the subsequent stability analysis \citep{Law1983, Hager1985, AB2008, Fog2012}. Their practical importance consists of supporting the design of stilling basins and similar hydraulic structures. The radial hydraulic jump, both as it is diverging or converging, is studied by \cite{VC2011, VC2013}, including the spatial development along its length.

The issue is also topical from an interdisciplinary viewpoint: recent experimental studies on physical hydraulic models are devoted to study the phenomenon of the standing accretion shock instability of collapsing stellar cores in astrophysics \citep{Fog2012}.

This work makes available reference analytical solutions. Recent work in the field of computational hydraulics has shown the importance of these analytical solutions \citep{Del2013} to validate the consistency, accuracy and robustness of shock-capturing numerical methods for the 1D and 2D SWE.

\section{The mechanical scheme} \label{sec:mech}
\begin{figure}
\begin{center}
\includegraphics[width=1.0\textwidth]{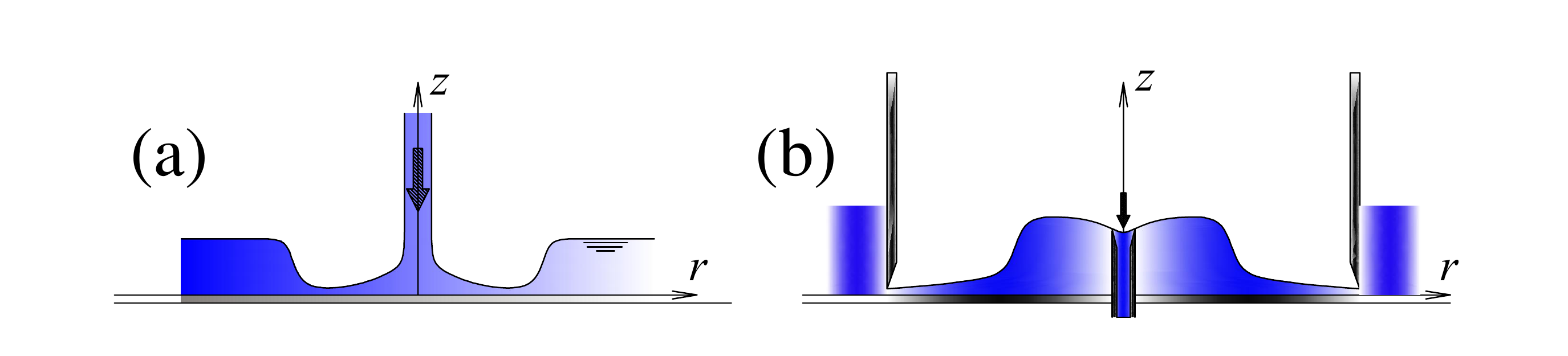}
\end{center}
\caption{Sketch of typical axially symmetric flows, in (a) centrifugal and (b) centripetal direction.}
\label{fig:disegno}
\end{figure}

An axially symmetric free surface radial flow is considered, due to the incidence of a vertical free jet against a plane horizontal plate, perpendicular to the jet axis, which is designated as $z$ axis. The radial velocity is positive if directed outward from the centre of the coordinate system $(r, \theta, z)$. Sufficiently far from the $z$ axis, the vertical velocity is small with respect to the radial velocity, and the tangential velocity is null. The same flow structure is valid, except for the sign of $u$, for an axially symmetric radial centripetal flow, which flows from the external boundary of a circular plate towards its centre. The feeding (from outside) results from a circular sluice gate, along with a free fall into a central pipe at the plate centre. The flow moves downward in the central free flowing weir. Figure~\ref{fig:disegno} provides a sketch of the reference flow field.
All classical hypotheses of the SWE are assumed applied to an incompressible liquid in the gravitational field. Inertial and gravity effects are considered as dominant with respect to viscous effects. Surface tension is neglected. The pressure distribution over each vertical is hydrostatic; the radial velocity is assumed uniform on each vertical; and the vertical velocity is assumed negligible with respect to the radial velocity. The axial symmetry is maintained everywhere.

\subsection{Basic steady flow equations}
Under the specified assumptions, the continuity and dynamic equations for radial steady flow read
\begin{equation} \label{eq:contdyn}
\frac{\partial}{\partial r}\left(U\,r\,Y\right)=0 \, ; \qquad U\frac{\partial U}{\partial r}+
g\frac{\partial Y}{\partial r}+g\frac{\partial z_b}{\partial r}+\frac{f}{2}\frac{U^2}{Y}=0
\end{equation}
Here $Y$ is the flow depth; $U$ the vertically-averaged radial velocity ($r$ direction); $g$ the gravity acceleration; $z_b$ the bottom elevation; $f$ the friction coefficient defined by $\tau_{0}=\left({1/2}\right)\,f\,\rho\,U^2$;  $\rho$ the liquid density, assumed constant; and $\tau_0$ the bed shear stress.
A reference liquid discharge, $Q$, is considered, which flows in a reference circular sector of half angular amplitude $\alpha$ ($\alpha=\pi$ in the fully circular case). $\,Q/\left({2} \, \alpha\right)$  is the liquid discharge per unit angular width. A reference specific energy, $E_{0}$, is also considered, which is related to the steady inviscid flow over the flat bed, with $E=Y+Q^{2}/\left[{2} g  \left(2 \alpha r Y \right)^{2}\right]$.
The total force of the flow is $F=\left({1/2}\right)\rho g \, r Y^{2} + \rho Q^{2}/ \left(2 \alpha r Y \right)$. The reference steady flow is characterised by a constant specific energy $E_{0}$ and a constant liquid discharge $Q$.

Non-dimensional equations are derived from Eq. (\ref{eq:contdyn}) with the critical depth as the vertical length scale, $Y_{0}=Y_c=\left({2/3}\right)E_0$; the critical radius $r_{0}=r_c=\left[Q/\left(2\alpha Y_c\sqrt{g\,Y_c}\right)\right]$ as the longitudinal length scale; and the critical velocity,  $U_{0}=U_c=\sqrt{g Y_c}$ as velocity scale. Critical quantities $\,Y_c\,$ and $\,r_c\,$ are defined as those minimising the specific energy and the total force, respectively \citep{VC2011}.
The non-dimensional reference discharge is $\varGamma=\left[Q/\left(2\,\alpha\,E_0^2\,\sqrt{gE_0}\right)\right]$.
It follows that the typical aspect ratio of the problem is $\beta=r_c/Y_c=\left({3/2}\right)^{5/2}\varGamma\simeq {2.76}\varGamma$.
The non-dimensional radius, depth, velocity, bottom elevation are, respectively: $\xi=r/r_0$; $\eta=Y/Y_0$; $u=U/U_0$; $\zeta=z_b/Y_0$.

For steady flow the continuity equation is reduced to the condition of constant liquid discharge as
\begin{equation} \label{eq:adcontinuity}
\left(u\,\xi\,\eta\right)=1
\end{equation}
The dynamic equation is expressed in terms of $\mathcal{F}=\left[\left({1/2}\right)\,\xi\,\eta^2+{1}/\left(\xi\,\eta\right)\right]$, the non-dimensional total force, or in terms of $\mathcal{E}=\left[\eta+{1}/\left(2\,\xi^2\,\eta^2\right)\right]$, the non-dimensional specific energy, with $\mathcal{H}=\zeta+\mathcal{E}$ as the non-dimensional total head
\begin{equation} \label{eq:adforce}
\frac{d}{d\xi}\left(\frac{1}{2}\,\xi\,\eta^2+\frac{1}{\xi\,\eta}\right)=
\frac{1}{2}\,\eta^2-\frac{1}{2}\,\beta\,f\,\left(\frac{1}{\xi\,\eta^2}\right)-\,\xi\,\eta\,\frac{d\zeta}{d\xi}
\end{equation}
\begin{equation} \label{eq:adenergy}
\frac{d}{d\xi}\left(\zeta+\eta+\frac{1}{2\,\xi^2\,\eta^2}\right)=-\frac{1}{2}\,\beta\,f\,\left(\frac{1}{\xi^2\,\eta^3}\right)
\end{equation}
Both formulations, which are equivalent for a continuous solution (though not for a discontinuous solution, as well known from the shallow-water theory of inviscid shocks) are useful to determine analytical solutions and to discuss a detailed analysis of the conservation properties of the system.

\section{Analytical solution: inviscid flow over flat horizontal bed} \label{sec:inviscid flat}	
An analytical solution is obtained by setting $\zeta={0}$ and $f={0}$ in Eq. (\ref{eq:adenergy}) and corresponds to the specific energy conservation $E=E_{0}=\left({3/2}\right)Y_c$, $\mathcal{E}={3/2}$, in the entire flow domain. The solution for the flow depth is apparent in the following implicit form
\begin{equation} \label{eq:csi}
\xi=\frac{1}{\eta\sqrt{3-2\,\eta}}
\end{equation}
This relationship determines the position where a prescribed depth occurs. Equation (\ref{eq:csi}) can be inverted using symbolic software (i.e., Mathematica, see www.wolframalpha.com) to obtain three solutions in the complex field. A procedure, omitted here for brevity, which bears similarities to that in \cite{VC2008} is finally used. 

\begin{figure}
\begin{center}
\includegraphics[width=0.48\textwidth]{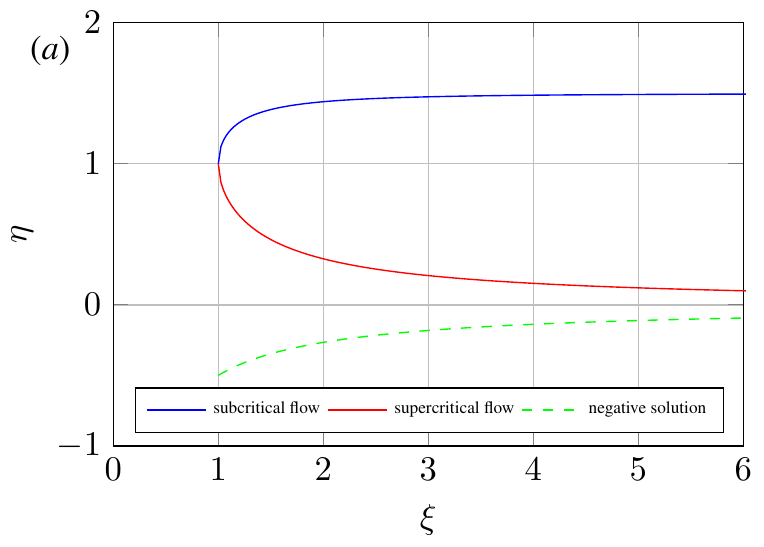}
\includegraphics[width=0.48\textwidth]{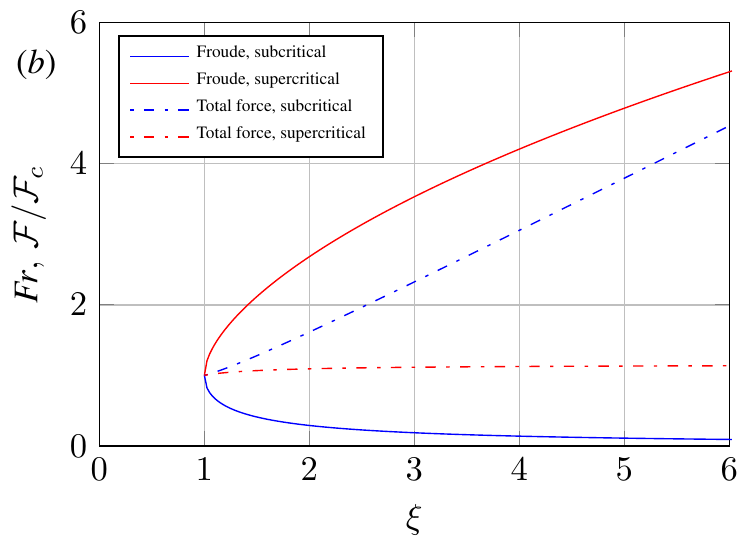}
\end{center}
\caption{Flow depth profiles in inviscid steady flow over flat horizontal bed;
(\textit{a}) depth $\eta\left(\xi\right)$ \textit{vs.} radius $\xi$,
(\textit{b}) Froude number $\Fr$ and $\mathcal{F}/\mathcal{F}_c$ \textit{vs.} radius $\xi$}
\label{fig:inviscid_flat}
\end{figure}

In the flow domain of physical interest ($\xi\geq {1};\,\,\varphi=\arcsin\left({1}/{\xi}\right);\,\,\pi/{2}\geq\varphi>{0}$), the solutions read
\begin{equation} \label{eq:real profiles}
\begin{aligned}
\eta_{1}&=\eta_{sb}=\frac{1}{2}+\cos\left(\frac{2}{3}\,\varphi\right) &: {1}\leq\eta_{sb}<\frac{3}{2}\\
\eta_{2}&=\eta_{sp}=\frac{1}{2}-\frac{1}{2}\,\cos\left(\frac{2}{3}\,\varphi\right)+\frac{\sqrt{3}}{2}\,
\sin\left(\frac{2}{3}\,\varphi\right) &: {1}\geq\eta_{sp}>{0}
\end{aligned}
\end{equation}
These solutions represent the explicit analytical solutions, not yet provided in the literature, namely: the subcritical (\textit{sb}) flow solution and the supercritical (\textit{sp}) flow solution. The physically meaningless, negative depth solution, is omitted here but is shown in Fig.~\ref{fig:inviscid_flat}. The trends of the Froude number $\Fr=u\,\eta^{-1/2}$, and of the total force $\mathcal{F}$ are also plotted.

\section{Analytical solution: flat horizontal bed with friction} \label{sec:friction flat}
This solution is obtained by setting $\zeta ={0}$ in Eq. (\ref{eq:adenergy}) and by referring to the specific energy $E=E_{0}=\left({3/2}\right)\,Y_c\,$, $\mathcal{E}=\left({3/2}\right)$, in the critical section. The friction term is computed by assuming a constant (small) friction coefficient. Equations (\ref{eq:csi}) and (\ref{eq:real profiles}), with reference only to the physical meaningful solutions, are considered to be the basic $0^{th}$-order solutions inside a perturbation approach as
\begin{equation} \label{eq:perturb}
\eta=\eta_{0}+\epsilon\,\eta_{1}+\ldots\, ; \qquad \,\epsilon=\frac{1}{2}\,\beta\,f\ll {1}
\end{equation}
At the $0^{th}$-order, the inviscid solution applies, determined in the previous section. It suffices to establish that $\eta_{0}=\eta_{sb}$ or $\eta_{0}=\eta_{sp}$ in Eqs. (\ref{eq:real profiles}), depending on the case.

It is straightforward to demonstrate that at first order, the depth profile is the solution of the differential equation
\begin{equation} \label{eq:eta1diff1}
\frac{d}{d\xi}\left(\eta_{1}-\frac{\eta_{1}}{\xi^{2}\,\eta_{0}^{3}}\right)=-\frac{1}{\xi^{2}\,\eta_{0}^{3}}
\end{equation}
Using mathematics, Eqs. (\ref{eq:csi}) and (\ref{eq:perturb}), the solution is found as
\begin{equation} 
\eta_{1}=\left[\frac{\left(\eta_{0}-{1}\right)\sqrt{{3}-{2}\,\eta_{0}}}{{2}\,\eta_{0}^{2}}+
\frac{\textrm{arctanh}\left(\sqrt{{1}-\left({2/3}\right)\eta_{0}}\right)}{\sqrt{3}} - \frac{\textrm{arctanh}\left({1}/\sqrt{3}\right)}{\sqrt{3}}\right]
\frac{\eta_{0}}{{3}\left(\eta_{0}-{1}\right)}
\label{eq:eta1}
\end{equation}
In Eq.  (\ref{eq:eta1}), \textrm{arctanh} is the hyperbolic arc tangent function. The integration constant is determined by establishing the critical depth at the critical radius, i.e.: $\eta_{1}({1})={0}$.
\begin{figure}
\begin{center}
\includegraphics[width=0.48\textwidth]{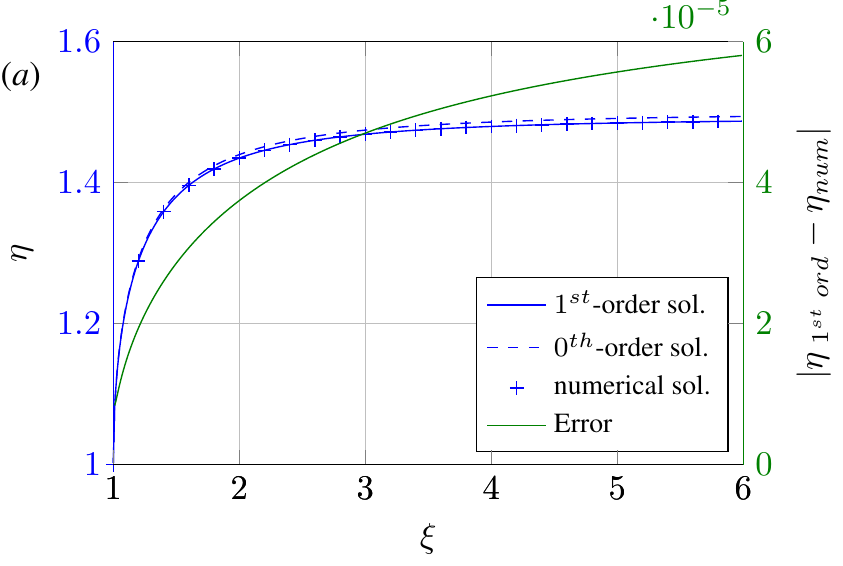}
\includegraphics[width=0.48\textwidth]{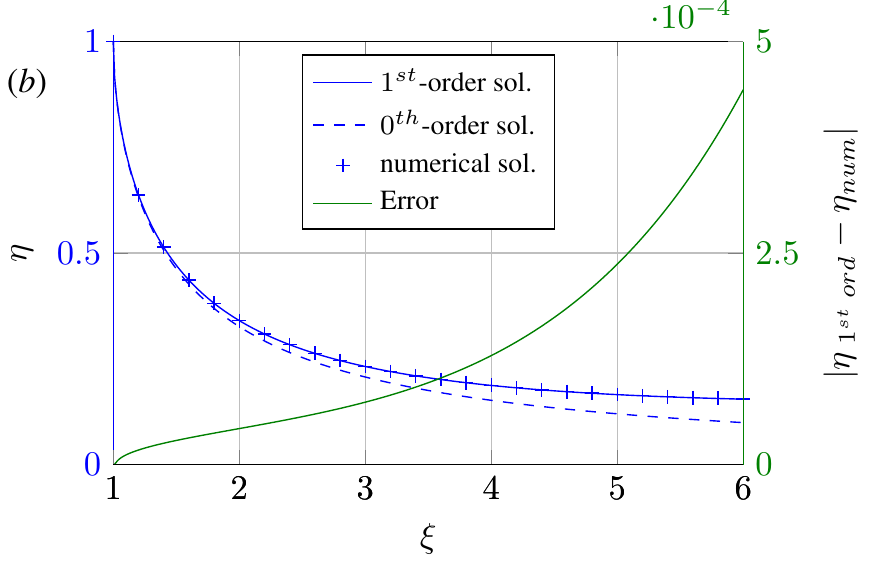}
\end{center}
\caption{Comparison between inviscid solution, numerical solution and analytical
first order solution for $\epsilon={0.02}$; (\textit{a}) subcritical flow, (\textit{b}) supercritical flow}
\label{fig:real_flat}
\end{figure}
The proposed analytical solution is valid for both cases, i. e. when the basic solution is super- or subcritical. Figure~\ref{fig:real_flat} compares both flow states, with the complete solution of Eq. (\ref{eq:adenergy}) determined with a Runge-Kutta 4th order numerical method \citep{BT1992}. For the selected values of the parameters, Fig.~\ref{fig:real_flat} shows the computed error, which is the difference between the analytical depth computed from Eqs. (\ref{eq:perturb}) and (\ref{eq:eta1}) and the numerical solution of Eq. (\ref{eq:adenergy}). The analytical solution is strictly close to the numerical, so that the perturbation procedure (\ref{eq:perturb}) \label{eq:} ends at first order.

\section{Analytical solution: inviscid flow over a variable bottom topography} \label{sec:inviscid variable bottom}
This solution is obtained by setting, in Eq. (\ref{eq:adenergy}), $\zeta\neq{0}$ and $f={0}$, and by considering that $\zeta =\zeta(\xi)$. The procedure to obtain the analytical solution is similar to that of \cite{VC2008}.
Consider as a constant the total head $H=z_b+E=\left({3/2}\right)Y_c$, and its non-dimensional counterpart, $\mathcal{H}=\zeta+\mathcal{E}=\left({3/2}\right)$. By setting $\psi={2}\left({3/2}-\zeta\right);\,\,\phi=\arcsin\left[{3}\,\sqrt{3}/\left(\psi^{3/2}\,\xi\right)\right]\,$ in the flow domain of physical interest (discarding the negative depth solution) the solutions are
\begin{equation}
\begin{aligned}
\eta_{sb}&=\frac{1}{6}\,\psi+\frac{1}{3}\,\psi\,\cos\left(\frac{2}{3}\,\phi\right) \\
\eta_{sp}&=\frac{1}{6}\,\psi-\frac{1}{6}\,\psi\,\cos\left(\frac{2}{3}\,\phi\right)+
                                  \frac{\sqrt{3}}{6}\,\psi\,\sin\left(\frac{2}{3}\,\phi\right) \label{eq:real z-profiles}
\end{aligned}
\end{equation}

These solutions are the explicit analytical solutions for flow depth: the subcritical solution $\eta_{sb}=\eta_{sb}(\xi)$ and the supercritical solution $\eta_{sp}=\eta_{sp}(\xi)$, respectively. They satisfy the implicit relationship $\xi=\left[{1}/\left(\eta\sqrt{\psi-{2}\,\eta}\right)\right]$.

The condition to obtain a real solution is
\begin{equation} \label{eq:z condition}
\frac{{3}\sqrt{3}}{\psi^{3/2}\,\xi}\leq{1}\quad\Rightarrow\quad
\zeta\leq\frac{3}{2}-\frac{1}{2}\left(\frac{{3}\sqrt{3}}{\xi}\right)^{2/3}
\end{equation}
thus indicating that the bottom elevation must be sufficiently small with respect to the available total head; the threshold also depends on the non-dimensional position. Flow choking requires the treatment of the hydraulic jump, which is analysed separately in the following section. In other words, a hydraulic jump occurs if inequality (\ref{eq:z condition}) is unsatisfied.
\begin{figure}
\begin{center}
\includegraphics[width=0.48\textwidth]{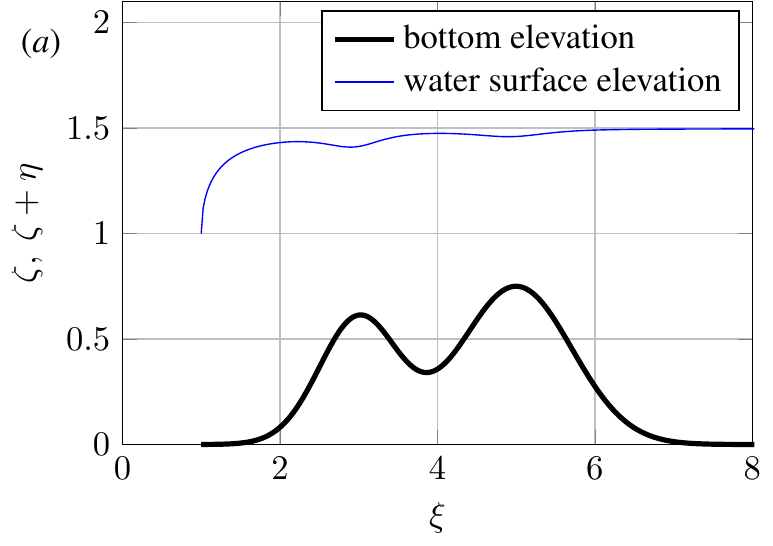}
\includegraphics[width=0.48\textwidth]{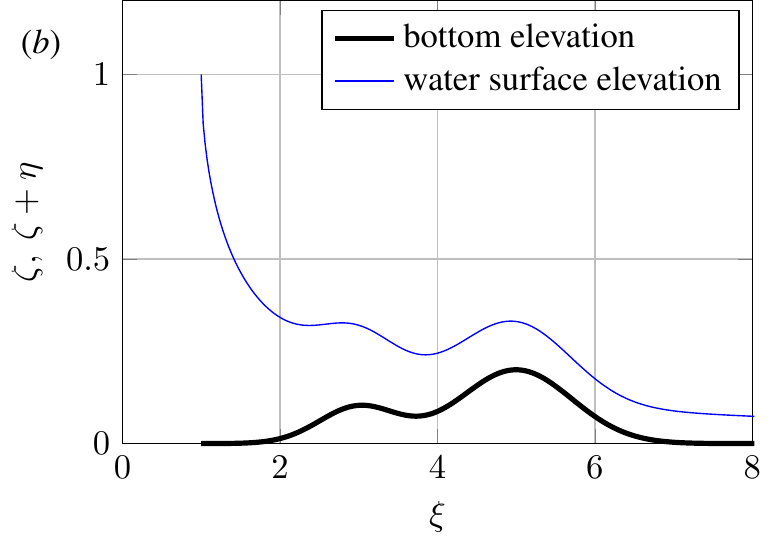}
\end{center}
\caption{Inviscid flow over uneven bottom;
(\textit{a}) subcritical flow, (\textit{b}) supercritical flow}
\label{fig:inviscid_uneven}
\end{figure}
In Fig.~\ref{fig:inviscid_uneven}, the behaviour of the free surface (without choking) is shown for sub- and supercritical flows, respectively. The bottom elevation is assumed as given according to the following equations, which are applied, respectively, for the sub- and supercritical cases
$\,\zeta=\left({3/5}\right) \exp{\left[-{2}\left(\xi-{3}\right)^{2}\right]}+\left({3/5}\right) \exp{\left[-\left(\xi-{5}\right)^{2}\right]}$; $\,\zeta=\left({1/10}\right) \exp{\left[-{2}\left(\xi-{3}\right)^{2}\right]}+\left({3/5}\right) \exp{\left[-\left(\xi-{5}\right)^{2}\right]}$.

\section{Analytical solution: direct hydraulic jump} \label{sec:hydraulic jump}
To ensure simplicity and generality, this solution is determined for the simplest conditions, described above (flat bottom, inviscid flow). The jump is considered an inviscid shock of zero-length, including the entire energy dissipation.
A more detailed treatment, which incorporates the gradual variability of the physical quantities inside the jump, which is inversely considered to have a finite length, is reported by \cite{VC2011} and \cite{VC2013}.  Here, the scheme concerning the mechanics is much more simple and general, even if less detailed. Neglecting the length of the jump allows to obtain an analytical solution for sequent depths and jump position, which cannot be obtained previously.
In the inviscid frame, the specific energy $E_{1}$ is a constant upstream of the jump, whereas a different, lower constant value $E_{2}$ corresponds to the downstream flow portion. All quantities are made non-dimensional using the upstream depth and upstream radius. Note that the word 'upstream' is used in the physical sense of the flow direction, so that it indicates smaller values of the radius for diverging flows and larger values of the radius for converging flows.

Using these hypotheses \cite{VC2011} demonstrate that the only non-dimensional parameter governing the phenomenon is the specific energy ratio $\mathcal{E}_R=E_{2}/E_{1}$, whose completion to unity $({1}-\mathcal{E}_R)$ is the rate of mechanical energy per unit weight dissipated in the jump.
The implicit expressions for the supercritical and subcritical free surface profiles are $\,\, \xi=\left[{1}/\left(\eta_{sp}\sqrt{{3}-{2}\,\eta_{sp}}\right)\right]$; $\xi=\left[{1}/\left(\eta_{sb}\sqrt{{3}\,\mathcal{E}_R-{2}\,\eta_{sb}}\right)\right]$.

Let superscripts $*$ and $**$ be the sequent quantities upstream and downstream of the jump, respectively (the depth and the velocity giving the same total force). Denoting $\xi_j$ as the shock position, the conditions at the jump are: \textit{i)} the uniqueness of jump position, \textit{ii)} mass conservation, and \textit{iii)} total force conservation as $\, \xi_{j}=\xi^{\ast}=\xi^{\ast\ast}\,$; $\, \left(u\,\eta\right)^{\ast}=\left(u\,\eta\right)^{\ast\ast}\,$; $\, \left[\left({1/2}\right)\,\xi\,\eta^{2}+u^{2}\,\eta\right]^{\ast}=\left[\left({1/2}\right)\,\xi\,\eta^{2}+u^{2}\,\eta\right]^{\ast\ast}\,$.
A nonlinear system (with $\eta^\ast$ and $\eta^{\ast\ast}$ as unknowns) is obtained.
The fundamental equation for the sequent depth ratio, $\Lambda=\eta^{\ast\ast}/\eta^{\ast}$ then is
\begin{equation} \label{eq:jump equation}
\mathcal{E}_R\,\Lambda^{3}-\left(4-\mathcal{E}_R\right)\,\Lambda^{2}+\left(4\,\mathcal{E}_R-{1}\right)\,\Lambda-{1}={0}
\end{equation}
In the range ${0}<\mathcal{E}_R<{1}$, it has only one real solution
\begin{equation} \label{eq:con depths ratio}
\Lambda=\frac{{4}-\mathcal{E}_R}{{3}\,\mathcal{E}_R}+\frac{\Upsilon}{{3}\,\mathcal{E}_R}+
\frac{{16}-{5}\,\mathcal{E}_R-{11}\,\mathcal{E}_R^{2}}{{3}\,\Upsilon\,\mathcal{E}_R}
\end{equation}
with: $\Upsilon=\left({64}-{30}\mathcal{E}_R-{51}\mathcal{E}_R^{2}+{17}\mathcal{E}_R^{3}+
{9}\sqrt{{20}\mathcal{E}_R^{2}+\mathcal{E}_R^{3}-{42}\mathcal{E}_R^{4}+\mathcal{E}_R^{5}+{20}\,\mathcal{E}_R^{6}}\right)^{1/3}$.

The corresponding super- and subcritical depths are
\begin{equation} \label{eq:con depths}
\eta^\ast=\frac{6}{\Lambda^{2}+\Lambda+{4}}\,;\quad
\eta^{\ast\ast}=\frac{{6}\,\Lambda}{\Lambda^{2}+\Lambda+{4}}
\end{equation}
The shock position is
\begin{equation} \label{eq:jump position}
\xi_j=\xi_j^\ast=\frac{1}{\eta^\ast\sqrt{{3}-{2}\,\eta^\ast\,}}=\xi_j^{\ast\ast}=
\frac{1}{\eta^{\ast\ast}\sqrt{{3}\,\mathcal{E}_R-{2}\,\eta^{\ast\ast}\,}}
\end{equation}
Usually, the literature gives the sequent depth ratio, the sequent depths and the jump position as functions of the upstream Froude number $\Fr^{\ast}$ (or some equivalent quantity), and more or less complicate computations are required to find its position. Note that Eqs. (\ref{eq:con depths ratio}), (\ref{eq:con depths}), (\ref{eq:jump position}) are really predictive, at least in the framework of the inviscid-shock theory. For a prescribed value of discharge, the quantity $\mathcal{E}_R$ is directly found if well-posed boundary conditions are known.
\begin{figure}
\begin{center}
\includegraphics[width=0.48\textwidth]{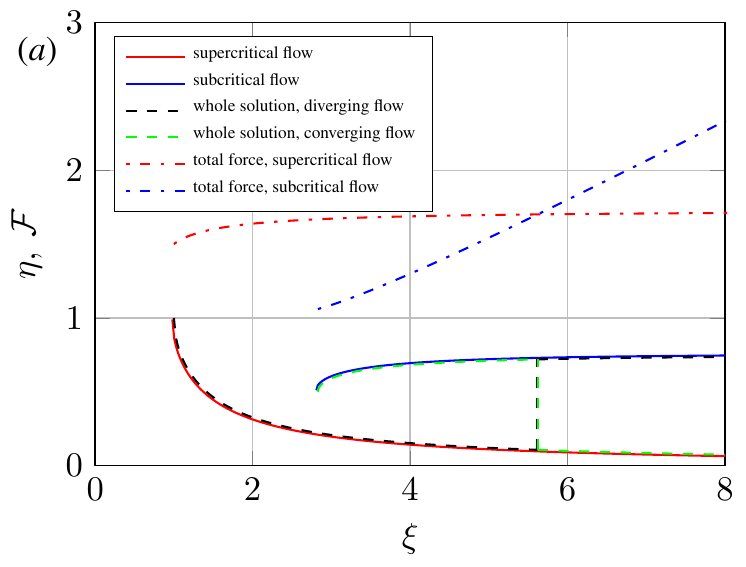}
\includegraphics[width=0.48\textwidth]{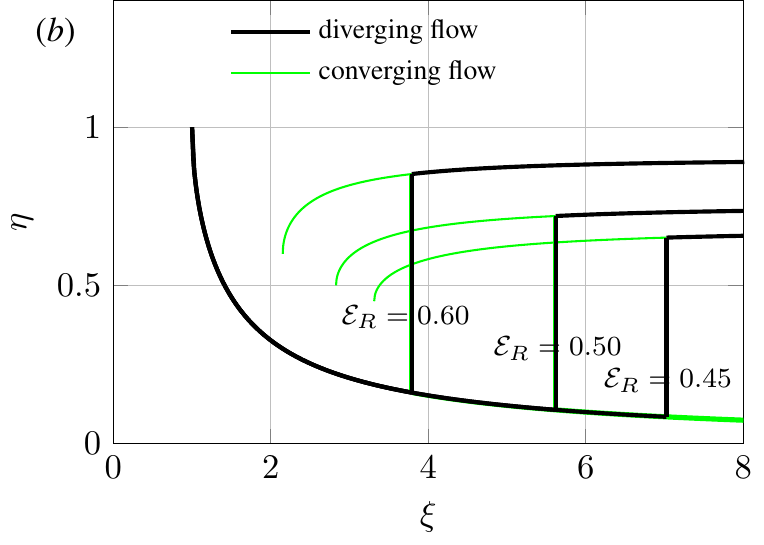}
\end{center}
\caption{Hydraulic jump
(\textit{a}) depth and force profiles $\eta\left(\xi\right)$ and $\mathcal{F}\left(\xi\right)$ for $\mathcal{E}_R={0.5}$, (\textit{b}) depth profiles $\eta\left(\xi\right)$ for different $\mathcal{E}_R$ values for diverging and converging flows}.
\label{fig:jump}
\end{figure}
\begin{figure}
\begin{center}
\includegraphics[width=0.48\textwidth]{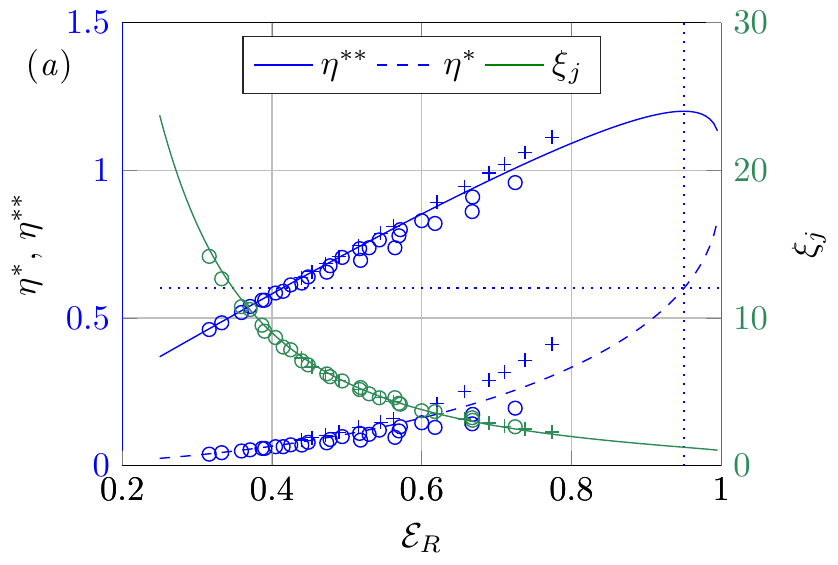}
\includegraphics[width=0.48\textwidth]{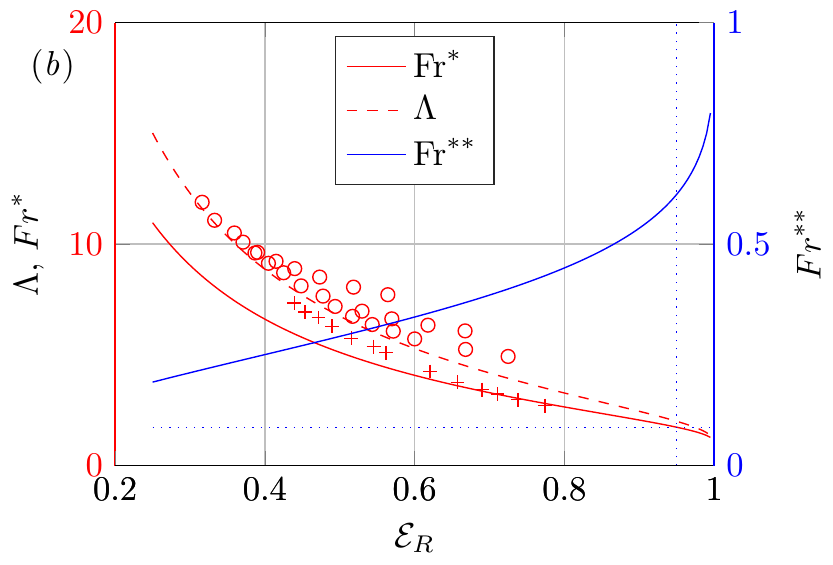}
\end{center}
\caption{Hydraulic jump
(\textit{a}) Sequent depths and jump position \textit{vs.} energy ratio, (\textit{b}) Sequent depths ratio and sequent Froude numbers \textit{vs.} energy ratio. (...) limit direct/undular jump. Experimental data from \citep{Rub63, Rub64}: (+) diverging, (o) converging; experimental points in (\textit{b}) refer to sequent depths ratio} 
\label{fig:condepths}
\end{figure}
In Fig.~\ref{fig:jump}, for a prescribed value of the energy ratio $\mathcal{E}_R={0.5}$, the behaviour of a jump is plotted (\textit{a}) using the derived equations, whereas the physical flow features are plotted in (\textit{b}) for three different values of $\mathcal{E}_R$.

Figure~\ref{fig:condepths} shows (\textit{a}) the behaviour of the sequent depths and the jump position versus the energy ratio and (\textit{b}) the behaviour of the sequent depths ratio and sequent Froude numbers, as functions of the same parameter. A comparison with experimental data from \cite{Rub63} and \cite{Rub64} is also shown. The selected dataset is chosen because experiments were performed at a quite large scale; further details are given in \cite{VC2011} and \cite{VC2013}. The dashed lines represent existence limits for the direct jump. Under the classical limit ${Fr}^{\ast 2}<{3}$,  corresponding to $\eta^{\ast}>{0.6}$  and $\mathcal{E}_R>{0.95}$, the undular jump occurs, for which the present theory does not hold. Notably, the above-mentioned limit for the upstream Froude number matches the condition that the undular jump occurs when no greater than five percent of the available specific energy is dissipated in the jump.

\section{Conclusions} \label{sec:Conclusions}
Analytical results concerning radial, axially symmetric, steady free surface flows are determined and discussed. The selected results can be used in field-scale hydraulic engineering because they pertains to the radial flow in stilling basins, where gravitational and inertial effects are dominant. The simplest case is the explicit solution for a flat horizontal bottom under inviscid flow. Analytical expressions for the sub- and supercritical flow depths are determined. An additional solution in the form of a perturbation solution is presented for the flat bottom and frictional flows, with the limit of small friction, both for sub- and supercritical flows. An analytical solution is also determined for inviscid flow over a spatially-varied bottom elevation, both for sub- and supercritical flows. The existence condition for these types of flows is determined and consists of the hypothesis of sufficiently small bottom elevation with respect to the prescribed level of the total head.

Analytical expressions are determined for the sequent depths and the jump position over a flat bed as functions only of the prescribed energy dissipation rate. This quantity is the unique parameter governing the phenomenon for inviscid flow outside the jump and based on the hypothesis of an inviscid shock of zero length.
These analytical results represent useful benchmarks to test numerical integration schemes for Shallow-Water Equations and as important reference conditions for the stability analysis of the radial hydraulic jump.

\pagebreak
%
%
%
\appendix\label{section:references}
%
%

%
%
%
\section{Notation}
\emph{The following symbols are used in this paper:}
\nopagebreak
\par
\begin{tabular}{r  @{\hspace{1em}=\hspace{1em}}  l}
$E, \mathcal{E}$ 	& specific energy of the flow (m), non-dimensional specific energy (-); \\
$\mathcal{E}_R$	& specific energy ratio between flows downstream/upstream the hydraulic jump (-); \\
$F, \mathcal{F}$ 	& total force of the flow (N), non dimensional total force (-); \\
$\Fr$ 			& Froude number (-); \\
$f$ 			& friction coefficient (-); \\
$g$			& gravity acceleration (m\,s$^{-2}$); \\
$H, \mathcal{H}$	& total head of the flow (m), non-dimensional total head (-); \\
$Q$			& liquid discharge (m$^3$\,s$^{-1}$); \\
$(r, \theta, z)$	& radial, angular, vertical coordinate (m, -, m); \\
$U, u$		& vertically averaged radial velocity (m\,s$^{-1}$), non dimensional radial velocity (-); \\
$Y, \eta$		& flow depth (m), non-dimensional flow depth (-); \\
$z_b, \zeta$		& bottom elevation (m), non-dimensional bottom elevation (-); \\
$\alpha$		& half angular amplitude of the reference circular sector (-); \\
$\beta$		& aspect ratio of the flow (-); \\
$\varGamma$	& non-dimensional reference discharge (-); \\
$\epsilon$		& non-dimensional small parameter in equation (\ref{eq:perturb}) (-); \\
$\Lambda$		& sequent depths ratio in the hydraulic jump (-); \\
$\xi$			& non-dimensional radial coordinate (-); \\
$\rho$		& liquid density (kg\,m$^{-3}$); \\
$\tau_{0}$		& bed shear stress (N\,m$^{-2}$); \\
\end{tabular}
\end{document}